\documentstyle[12pt]{article}
\begin{document}

\vspace*{0.5cm}
\baselineskip=0.8cm

\vspace*{0.5cm}
\baselineskip=0.8cm
\begin{center}
{\Large\bf Temperature Dependent Suppression  of Conductance 
in Quantum Wires: Anomalous Activation Energy from Pinning of
the Band Edge}
\vspace{2.0cm}

{\large\bf Kenji Hirose$^{1}$ and Ned S. Wingreen$^2$}
\vspace{0.7cm}

$^{1}$ Fundamental Research Laboratories, NEC Corporation, 34
Miyukigaoka, Tsukuba, Ibaraki 305-8501, Japan\\
$^{2}$ NEC Research Institute, 4 Independence Way, Princeton, New Jersey
08540\\
\vspace{0.5cm}
\end{center}

\baselineskip=0.72cm

\vspace{1.5cm}

\begin{center}
{\large\bf Abstract}
\end{center}

\vspace{0.6cm}

For unpolarized electrons in a clean quantum wire,
density functional theory reveals a thermally
activated suppression of conductance.
The activation temperature $T_A$ grows slowly
(roughly quadratically) with gate voltage due to pinning
of the band edge at the chemical potential.
Similar results were obtained experimentally 
by Kristensen {\it et al.} [Phys.~Rev.  {\bf B62}, 10950 (2000)]
for the temperature dependence of the anomalous
conductance plateau near 0.7 $(2e^2/h)$ in a
quantum point contact.

\vspace{1.0cm}
\noindent

\newpage

The discovery of the quantization of the conductance through a point 
contact in steps of $2e^2/h$ \cite{vanWees,Wharam} was a milestone
in mesoscopic physics. Recent observations of 
additional plateaulike structures at $0.7 (2e^2/h)$ in
quantum point contacts \cite{Thomas,Kristensen} 
and at $0.5 (2e^2/h)$ in clean quantum wires \cite{Reilly}
have been the center of much attention. 
These plateaulike structures become more pronounced 
as the temperature is raised. In the quantum point contacts, 
the conductance is 
suppressed by temperature between $0.7 (2e^2/h)$  and the 
plateau value $2e^2/h$, while very little
temperature dependence is observed below $0.7 (2e^2/h)$. 
Kristensen and co-workers \cite{Kristensen} found that the 
conductance suppression has an activated
behavior, with the activation temperature growing roughly 
quadratically with gate voltage.

The present paper reports density-functional-theory (DFT) results
for the thermally activated suppression 
of conductance for unpolarized electrons in a clean
quantum wire. As in the experiment \cite{Kristensen}, we
find an activation temperature $T_A$ that depends roughly
quadratically on side-gate voltage, $T_A \propto V_s^2$,
in the plateau region. 
We show that this anomalous behavior of the 
activation temperature derives from self-consistent
pinning of the band edge at the absolute electrochemical potential.

The electronic states of a clean quantum wire are obtained using 
density-functional theory within the local-density 
approximation \cite{Callaway}. This method allows us to study both 
confinement and electron-electron interactions in a unified
framework.  A number of works have considered the possibility
of spontaneous spin polarization of electrons in a quantum 
wire \cite{Wang96,Zabala98,Calmels98,Hirose}. 
Here we consider unpolarized electrons 
to highlight effects that are electronic but not magnetic 
in origin.  Exploiting translational invariance 
in the $x$ direction, we expand
the wavefunctions as $\Psi_{n,k_n}({\bf r})=e^{i k_n
x}\sum_{m}c_{n,m}\psi_{m}(y)$, where $n$ is the subband number and
$\psi_{m}(y)$ is the m${th}$ eigenvector of the bare parabolic
potential. The following Kohn-Sham equations \cite{Kohn} are solved
numerically, 
\begin{eqnarray}
&&\left[-\frac{\hbar^2}{2m^*}\nabla^2+\frac{1}{2}m^*\omega_y^2 y^2
-\frac{e^2}{\kappa}\int_{-\infty}^{\infty}\rho(y'){\rm ln}
\left[\frac{(y-y')^2}{(y-y')^2+a^2}\right]dy'
+\frac{\delta E_{\rm xc}[\rho]}{\delta\rho({\bf r})} \right]
\Psi_{n,k_n}({\bf r}) \nonumber \\
&&\hspace*{3.0cm}=(\epsilon_n+\frac{\hbar^2 k_n^2}{2m^*})
\Psi_{n,k_n}({\bf r}) \nonumber \\
&&\hspace*{2.0cm}\rho({\bf r})
=\frac{1}{\pi}\sum_{n}\int_{-\infty}^{\infty} f(\epsilon_n+\frac{\hbar^2
k_n^2}{2m^*})|\Psi_{n,k_n}({\bf r})|^2 dk_n,
\end{eqnarray}
for each electron density 
\begin{equation}
n_{1D}=\frac{1}{\pi}\sum_{n}\int_{-\infty}^{\infty}
f(\epsilon_n+\frac{\hbar^2 k_n^2}{2m^*}) dk_n. 
\label{noned}
\end{equation}
In the above equations, $\rho({\bf r})$ is the local electron
density, $\epsilon_n$ is the band-edge energy of the $n$th
subband, and the Fermi distribution function is 
$f(\epsilon)=1/[{\rm exp}\{(\epsilon-\mu_{\rm wire})/k_B T\}+1]$, 
where the electrochemical potential in the wire $\mu_{\rm wire}$ is
determined
self-consistently for each density $n_{1D}$ \cite{Hirose}. 

In the experiments \cite{Thomas,Kristensen,Reilly}, the side-gate
voltage 
$V_s$ is used to change the density $n_{1D}$. The absolute 
electrochemical
potential $\mu$ is the sum of the internal electrochemical 
potential of the wire $\mu_{\rm wire}$ and the external
potential energy. The latter varies   linearly with  side-gate
voltage as $-\alpha e V_s$, where the
capacitive lever-arm $\alpha$ 
depends on the detailed gate geometry \cite{Foxman}. To model the fixed
leads, 
we set the absolute electrochemical potential at $\mu = 0$, 
and take $\alpha=1$ for simplicity, so that 
$eV_s=\mu_{\rm wire}$. This also implies that at $T=0$
the lowest subband energy is set at $\epsilon_0 = 0$ 
when $V_s = 0$.

The conductance is calculated from the Landauer
formula,
\begin{equation}
G=\frac{2e^2}{h}\sum_{n}\int_{-\infty}^{\infty}\left(-\frac{\partial
f}{\partial \epsilon}\right)T_{n}(\epsilon) d\epsilon.
\label{landauer}
\end{equation}
To model a clean quantum wire, we assume 
that the transmission coefficients $T_{n}(\epsilon)$
change abruptly from 0 to 1 at the band-edge energies $\epsilon_n$.
We fix the external confining potential as
$\hbar\omega_y=2.0{\rm meV}$ 
and take an image-charge plane at $a = 100{\rm nm}$. The material
constants are those of GaAs, $m^*=0.067m$ and $\kappa=13.1$.

Figure 1(a) shows the conductance $G$ in the lowest subband as a
function 
of side-gate voltage $V_s$ at different temperatures T = 0.25K, 0.5K,
0.67K, 1.0K, and 2.0K, from top to bottom,
respectively \cite{instability}.
We see that the conductance is suppressed with increasing   
temperature. Guided by the experimental results of 
Kristensen {\it et al.} \cite{Kristensen}, we look for Arrhenius-type 
suppression of the conductance,
\begin{equation}
G(T) = G_0 (1 - e^{-T_A/T}),
\label{arrhenius}
\end{equation} 
where $G_0 = 2e^2/h$.
In the inset of Figure 1(b), we plot the relative
conductance
suppression as ${\rm ln}[1-G(T)/G_0]$ versus the inverse of the 
temperature at different side-gate voltages (1) $V_s=1.0{\rm mV}$, (2)
$V_s=3.0 {\rm mV}$ 
and (3) $V_s=5.0 {\rm mV}$.
The linear behavior of Arrhenius plots of this  
type allows us to extract activation temperatures 
$T_A=-T{\rm ln}[1-G(T)/G_0)]$ for all side-gate voltages $V_s$. 
Figure 1(b) shows that the activation temperature
has an approximately quadratic dependence on the side-gate voltage, 
close to that observed in the experiment of 
Kristensen {\it et al.} on quantum point contacts \cite{Kristensen}.

The roughly quadratic dependence of the activation temperature on
side-gate voltage in our DFT results reflects
``pinning'' of the band edge at the absolute electrochemical potential.
As electrons begin to enter the 
wire, their self-consistent potential buoys up
the band-edge energy.  The result is that the difference
between the absolute electrochemical potential and the band edge grows
slowly  with side-gate voltage. 

To explore the band-edge pinning effect, we use a simplified model
consisting of 
a purely 1D wire with a classical capacitance \cite{Hirose}. 
We find that this model captures the essential features
of the DFT results. The kinetic energy of the 1D electrons is 
$\epsilon_k=\hbar^2 k^2/2m^*$, and the band-edge 
energy $\epsilon_0 = -eV_s + (e^2/C)n_{1D}$ is the sum of 
the side-gate contribution and a classical capacitive energy,
where $C$ is a fixed capacitance per unit length of the wire. 
The electron density, for two spins, is determined by 
\begin{equation}
n_{1D}=\frac{1}{\pi}\int_{-\infty}^{\infty} \frac{dk}
{{\rm exp}\{(\epsilon_k- \mu_0)/k_B T\}+1}.
\label{density}
\end{equation}
For each $V_s$, the chemical potential $\mu_0$ of the 1D electrons is 
determined self-consistently by the condition that the  absolute
electrochemical potential $\mu$ is fixed at zero,
so that  $\mu =  \mu_0 + \epsilon_0 = \mu_0 - eV_s + (e^2/C)n_{1D} = 0$.
To remove the explicit dependence on the parameters $m^*$ and $C$, 
this condition can be written in dimensionless form as
\begin{equation}
\tilde{\mu}_0 - \tilde{V}_s + \int_{0}^{\infty}
\frac{d\tilde{k}}{{\rm
exp}\{(\tilde{k}^2-\tilde{\mu}_0)/\tilde{T}\} +1}=0.
\label{model}
\end{equation}
Here all the energies $\mu, k_B T$, and $eV_s$ are normalized by
$U=(e^2/C)^2/(\pi^2 \hbar^2/8m^*)$, and 
$\tilde{k} = \pi \hbar^2 k/[4 m^* (e^2/C)]$.

In the inset to Figure 2(b), we show the chemical potential
$\tilde{\mu}_0$ as a function of $\tilde{V}_s$, obtained
by solving Eq. (\ref{model}) for different temperatures.
In the zero temperature case, $\tilde{\mu}_0$ has a cusp where it
crosses zero, {\it i. e.} where electrons first enter the
wire. The relatively slow  increase
of $\tilde{\mu}_0$ with gate voltage on the right
side of the cusp reflects the pinning of the band
edge at the absolute electrochemical potential. 
To elucidate this effect, we consider the zero-temperature limit
of Eq. (\ref{model}),
\begin{equation}
\tilde{\mu}_0+\sqrt{\tilde{\mu}_0} -\tilde{V}_s=0,
\label{tzero}
\end{equation}
which has the solution 
\begin{equation}
\tilde{\mu}_0 = \frac{1}{2}\left[ 1 + 2 \tilde{V}_s
-\sqrt{1 + 4 \tilde{V}_s } \right] \simeq \tilde{V}_s^2, \ 
\tilde{V}_s \ll 1.
\label{tzerosol}
\end{equation}
The quadratic growth of the chemical potential
for side-gate voltage $\tilde{V}_s \ll 1$ 
can be traced to the square-root 
singularity $\rho_{1D}(\epsilon) \sim \epsilon^{-1/2}$ 
in the density of states of 1D electrons \cite{Hirose}: 
As electrons begin to enter the wire, because of this high
density of states, a very small increase in the chemical
potential $\mu_0$ causes a large increase in the density
of electrons $n_{1D}$. This, in turn, produces a  large increase in
the band-edge energy $\epsilon_0$ through the self-capacitance
of the wire $(e^2/C)n_{1D}$. The result is that a large
increase of $V_s$ is required for a small increase of $\mu_0$. 
As seen in the inset to Figure 2(b), the pinning effect
diminishes as temperature broadens the electron-energy 
distribution.

For the classical capacitance model, the conductance
can be obtained analytically from the Landauer formula
(\ref{landauer}) with a single subband,
\begin{equation}
G = G_0 f(0) = \frac{G_0}{\exp(-\tilde{\mu}_0/\tilde{T}) + 1},
\label{gclassical}
\end{equation}
where we have again assumed that the transmission coefficient
jumps abruptly from 0 to 1 at the band-edge energy $\epsilon_0$.
In Figure 2(a), the conductance is plotted as a function of 
side-gate voltage $\tilde{V}_s$ at different temperatures.
At $\tilde{T}=0$, the conductance $G$ abruptly  jumps from
$0$ to $2e^2/h$  at $\tilde{V}_s=0$, where the band edge drops below the
absolute electrochemical potential.
Increasing the temperature
suppresses the conductance with an approximately
activated form, $G(T) \simeq G_0 [1- \exp(-\tilde{T}_A/\tilde{T})]$.

The activation temperature $\tilde{T}_A$ for the suppression
of conductance can be obtained exactly and analytically as a function of
$\tilde{V}_s$ by considering
the low temperature limit of the conductance,
\begin{equation}
\tilde{T}_A \equiv -{\rm lim}_{\tilde{T}/\tilde{\mu}_0 \rightarrow 0} \
\ 
 \{ \tilde{T} \ln [1 - G(\tilde{T})/G_0]\} = \tilde{\mu}_0.
\label{activation}
\end{equation}
{\it The activation temperature is
just equal to the zero-temperature chemical potential.}
Hence, the initial quadratic behavior of the activation temperature
$\tilde{T}_A$ with gate voltage simply reflects the initial
quadratic behavior of $\tilde{\mu}_0$ with gate voltage due to the
pinning
of the band edge.  The activation temperature $\tilde{T}_A$
is plotted in Figure 2(b) using the full expression for $\tilde{\mu}_0$
from Eq. (\ref{tzerosol}). 

These results may shed light on the observation by
Kristensen {\it et al.} \cite{Kristensen} of thermally
activated suppression of conductance near the 
anomalous $0.7 (2e^2/h)$ conductance plateau in a 
quantum point contact (QPC). In a phenomenological
model, Bruus {\it et al.} \cite{Bruus} have attributed 
$T_A$, and its gate-voltage dependence, to proximity of the 
absolute electrochemical potential to a subband edge. 
Our observation of band-edge pinning
provides a microscopic justification 
for this model, insofar as a QPC can be regarded as
a segment of clean quantum wire.  
However, various aspects of transport through QPCs remain unaccounted
for microscopically, including the magnetic
field dependence \cite{Thomas}.

In conclusion, we have explored the thermally activated suppression of
conductance in a clean quantum wire, using both density
functional theory and a classical capacitance model for
unpolarized electrons.  
The quadratic dependence of the activation temperature on
the gate voltage in the plateau region 
reflects self-consistent pinning of the band
edge by the absolute electrochemical potential.  We hope that this work
contributes to the understanding of the thermally suppressed
conductance in  quantum point contacts and clean quantum wires. 
The authors acknowledge valuable conversations with Yigal Meir.

\newpage

\noindent
{\bf Figure Captions}
\begin{itemize}

\item[Figure 1:] 
(a) Conductance $G$ as a function of side-gate voltage $V_s$ at
different temperatures T = 0.25K, 0.5K, 0.67K, 1.0K, and 2.0K (from top
to bottom) 
within the DFT calculations with parameters appropriate to
GaAs, $\kappa=13.1$ and
$\hbar\omega_y=2.0{\rm meV}$.
(b)  Activation temperature $T_A$ as a function of $V_s$
obtained from Arrhenius  plots of conductance suppression.
Inset --  Arrhenius plots of conductance at fixed
side-gate voltages of $V_s=$ (1) $1.0 {\rm mV}$, (2) $3.0 {\rm mV}$,
and (3) $5.0 {\rm mV}$,
respectively.

\item[Figure 2:] (a) Conductances $G$ as a function of 
side-gate voltage $\tilde{V}_s$ at different temperatures 
from $\tilde{T}=0$ (above) to $\tilde{T}=0.1$ (below) in steps of
$\Delta\tilde{T}=0.01$ within the classical-capacitance model. 
(b) The activation temperature $\tilde{T}_A$ as a function of
$\tilde{V}_s$. Inset --
Chemical potential $\tilde{\mu}_0$ as a function of
$\tilde{V}_s$ 
at different temperatures from $\tilde{T}=0$ to $\tilde{T}=0.1$ (from
top to
bottom at $\tilde{V}_s=0$) in steps of $\Delta\tilde{T}=0.02$. 
Here $\tilde{V}_s$, $\tilde{T}$, $\tilde{T}_A$ and $\tilde{\mu}_0$ are
all measured in units of $U=(e^2/C)^2/(\pi^2 \hbar^2/8m^*)$.

\end{itemize}


\begin{thebibliography}{1000}
\bibitem{instability} Only part of the first conductance step is
shown because the uniform state is unstable at very low density
within density functional theory.
\end{thebibliography}
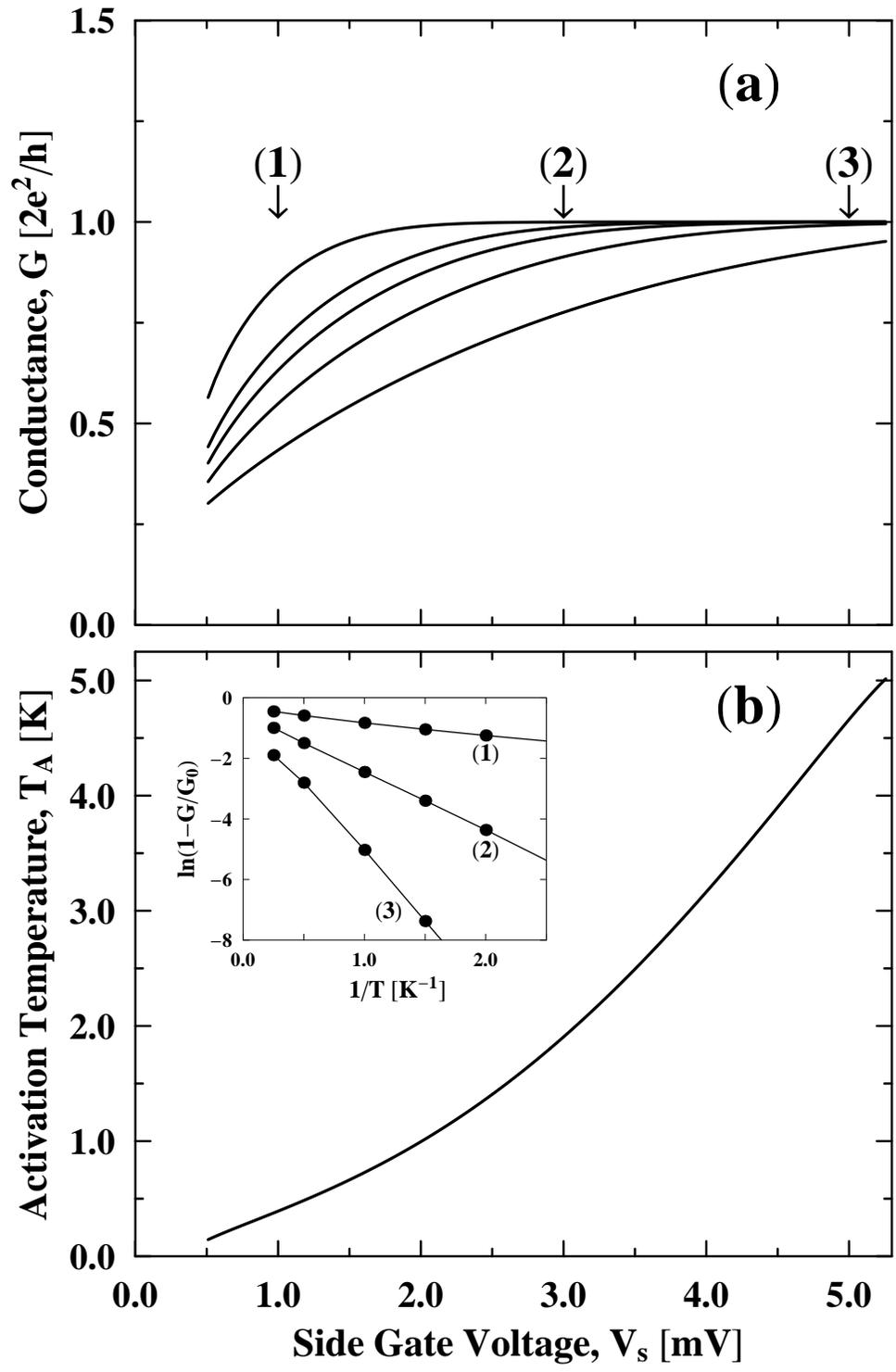
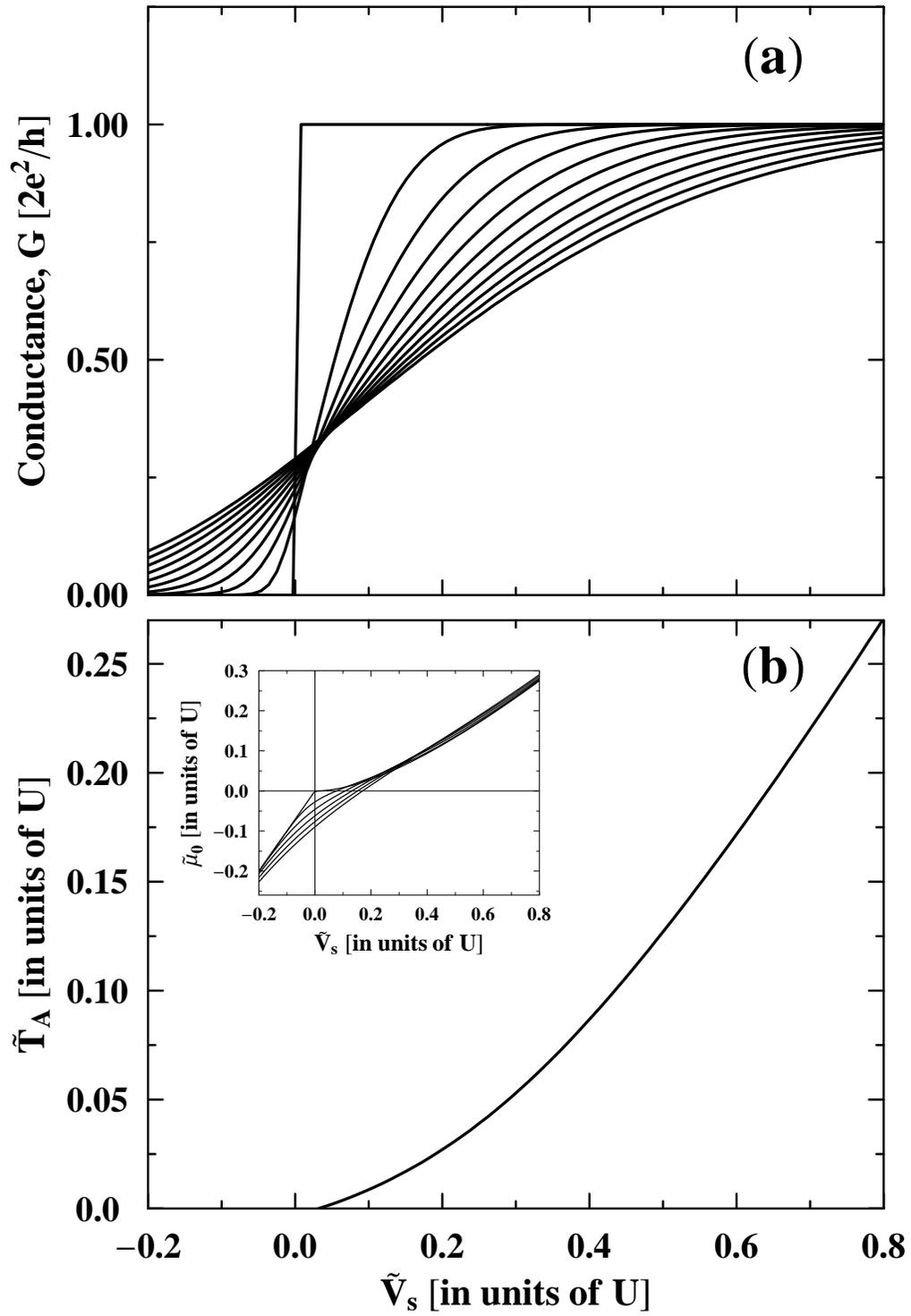
\end{document}